\documentstyle[aps,prb,amsfonts,twocolumn]{revtex}

\renewcommand{\vec}[1]{{\mathbf #1}}
\input psfig

\begin{document}

\title{Spatio-temporal speckle correlations for imaging in turbid
media$^1$}

\author{Serguei E. Skipetrov}

\address{Department of Physics, Moscow State University, 119899
Moscow, Russia.\\Webpage: http://skipetrov.chat.ru/, e-mail:
skipetrov@mail.ru}

\maketitle

\begin{abstract} \begin{center} \bf Abstract \end{center}
We discuss the far-field spatio-temporal cross-correlations of
waves mul\-ti\-ple-scattered in a turbid medium in which is embedded
a hidden heterogeneous region (inclusion)
characterized by a distinct scatterer
dynamics (as compared to the rest of the medium). We show that
the spatio-temporal correlation is affected by the inclusion
which suggests a new method of imaging in turbid media.
Our results allow qualitative interpretation in terms
of diffraction theory: the cross-correlation of scattered
waves behaves similarly to the intensity of a wave diffracted
by an aperture.
\end{abstract}
\vspace{1cm}

A considerable progress\footnotetext[1]{Published in {\em Waves and
Imaging Through Complex Media,} edited by P. Sebbah (Kluwer
Academic Publishers, Dordrecht, 2001).} has been made during the
recent years in the understanding of wave transport in disordered
media.\cite{rossum99} Very similar phenomena are shown to exist in
multiple scattering of electrons and classical waves (e.g., light)
under particular circumstances.\cite{lag96} Some of the concepts
developed first theoretically, and then studied in model
experiments, are now very close to practical applications. One of
the important fields where the physics of multiple-scattered waves
is currently finding its applications is the (medical) imaging of
disordered, turbid media.\cite{yodh95} The light waves scattered
inside a turbid medium (e.g., human tissue) carry information on
the properties of the medium. The information can be considered as
being ``encoded'' in the statistics of the waves. Analysis of the
latter statistics allows one to reconstruct (or ``image'') the
scattering medium.

\begin{figure}
\vspace{0.2cm} \psfig{file=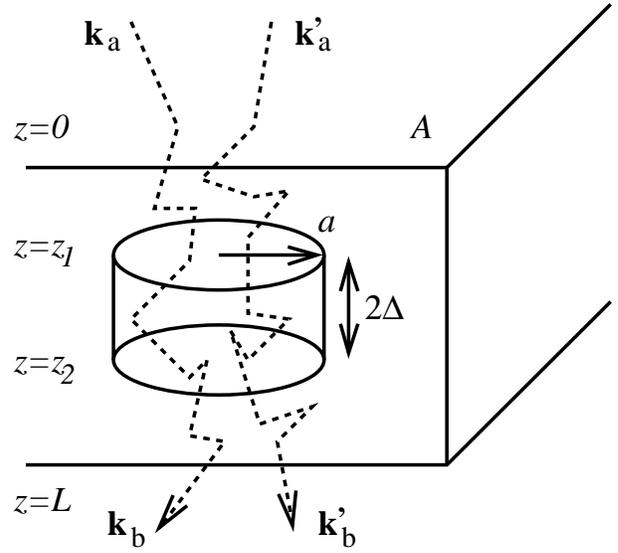,width=8cm} \vspace*{0.5cm}
\caption{A cylindrical inclusion of height $2\Delta = z_2 - z_1 \gg
\ell$ and radius $a \gg \ell$ ($\ell$ is a photon transport mean
free path) is embedded at $z_0 = (z_1 + z_2)/2$ inside a turbid
slab of width $L$ and surface area $A = W^2$, $W \gg L$, $W \gg a$.
$\vec{k}_a$ and $\vec{k}_a^{\prime}$ denote the wave vectors of
incident waves, while $\vec{k}_b$ and $\vec{k}_b^{\prime}$ --- the
wave vectors of transmitted waves. The scatterers in the medium
undergo Brownian motion with diffusion coefficients $D_{in}$
(inside the inclusion) and $D_{out}$ (outside the inclusion).}
\label{fig1}
\end{figure}

A simplified version of a typical geometry considered in connection
with imaging problems is shown in Fig.~\ref{fig1}. A slab of turbid
medium occupies the space between the planes $z=0$ and $z=L$, and
some region (a cylinder-shaped inclusion) inside the slab is
assumed to have somewhat different properties as compared to the
surrounding medium. If ``different properties'' means different
scattering $\mu_s^{\prime}$ and/or absorption $\mu_a$ coefficients,
one can image the inclusion by measuring the spatial distribution
of the average intensity $I(\vec{r}) = \left< E(\vec{r},
t)E^*(\vec{r}, t) \right>$ of transmitted (or reflected)
wave.\cite{oleary92,outer93} Here $E(\vec{r}, t)$ is the amplitude
of scattered wave at spatial position $\vec{r}$ at time $t$. If
$\mu_s^{\prime}$ and $\mu_a$ are constant throughout the medium,
and the contrast between the inclusion and surrounding medium is
provided by the scatterer dynamics (different types and/or
intensities of scatterer motion inside and outside the inclusion),
the methods of diffusing-wave spectroscopy \cite{maret87,pine88}
can be applied to visualize the inclusion.\cite{maret00} In the
latter case, one measures the time autocorrelation function
$C_1(\vec{r}, \tau) = \left<E(\vec{r}, t) E^*(\vec{r}, t+\tau)
\right>$ of scattered wave field at multiple positions $\vec{r}$,
which allows visualization of the inclusion.

In the present contribution, we propose to use the {\em
spatio-temporal cross-correlation\/} function $C_1(\vec{r},
\Delta\vec{r}, \tau) = \left<E(\vec{r}, t)
E^*(\vec{r}+\Delta\vec{r}, t+\tau) \right>$ for the purpose of
imaging in turbid media. Since the time autocorrelation function
$C_1(\vec{r}, \tau)$ carries more information about the turbid
medium than the average intensity $I(\vec{r})$, we suggest that the
information contents of the spatio-temporal correlation function
$C_1(\vec{r}, \Delta\vec{r}, \tau)$ should be even more rich. If
the points $\vec{r}$ and $\vec{r} + \Delta\vec{r}$ are taken far
enough from the medium (in the far-field of scattered wave),
$C_1(\vec{r}, \Delta\vec{r}, \tau)$ is equivalent to the
angular-temporal correlation function $\left<E(\vec{k}_b, t)
E^*(\vec{k}_b^{\prime}, t+\tau) \right>$ [where $E(\vec{k}, t)$ is
the spatial Fourier transform of $E(\vec{r}^{\prime}, t)$ with
$\vec{r}^{\prime}$ taken at the plane where the scattered waves
leave the medium].

We start with a macroscopically homogeneous turbid medium (no
inclusion), and assume that scatterers in the medium undergo
Brownian motion with a diffusion coefficient $D$. In addition, we
assume a weak-scattering limit $k\ell \gg 1$ (where $\ell =
1/\mu_s^{\prime}$ is the photon transport mean free path), and
neglect the absorption of light in the medium ($\mu_a = 0$). As
depicted in Fig.~\ref{fig1}, a plane wave is incident upon a turbid
slab at time $t$ with a wave vector $\vec{k}_a$. The transmitted
wave leaves the slab with a wave vector $\vec{k}_b$. Similarly, for
an incident wave with a wave vector $\vec{k}_a^{\prime}$ at time
$t+\tau$, the transmitted wave has a wave vector
$\vec{k}_b^{\prime}$. Assuming unit amplitudes of incident waves,
we calculate the correlation function of transmitted fields
$C_1(\vec{k}_a, \vec{k}_b; \vec{k}_a^{\prime}, \vec{k}_b^{\prime};
\tau) = \left< E(\vec{k}_a, \vec{k}_b, t) E^*(\vec{k}_a^{\prime},
\vec{k}_b^{\prime}, t+\tau) \right>$ using the standard
diagrammatic techniques,\cite{frisch68} following the general
calculation scheme developed in Ref.\ \onlinecite{berk94}. In the
leading order in a small parameter $1/k\ell$, we obtain:
\begin{eqnarray}
C_1 &=& \frac{\ell^2}{4 k^2 A^2} \int\int \mathrm{d}^2 \vec{R}_1 \;
\mathrm{d}^2 \vec{R}_2 \; \nonumber \\
&\times& \exp\left(-i \Delta\vec{q}_a \vec{R}_1 + i \Delta\vec{q}_b
\vec{R}_2 \right)
\nonumber \\
&\times&P\left( \left\{\vec{R}_1, \ell \right\}, \left\{ \vec{R}_2,
L-\ell\right\}, \tau \right),
\label{ladder}
\end{eqnarray}
where $\vec{q}$'s denote projections of $\vec{k}$'s onto the plane
$z=const$, $\Delta\vec{q}_a = \vec{q}_a - \vec{q}_a^{\prime}$,
$\Delta\vec{q}_b = \vec{q}_b - \vec{q}_b^{\prime}$, $\vec{r} =
\left\{ \vec{R}, z \right\}$ with $\vec{R}$ being a two-dimensional
vector perpendicular to the $z$-axis, and we assume the first and
the last scattering events to occur at $z=\ell$ and $z=L-\ell$,
respectively. If $L \gg \ell$ and $\left| \vec{r}_1 - \vec{r}_2
\right| \gg \ell$, the reduced ladder propagator $P$ entering into
Eq.\ (\ref{ladder}), obeys the diffusion equation:
\begin{eqnarray}
\left[ \nabla^2 - \alpha^2(\tau) \right] P(\vec{r}_1, \vec{r}_2,
\tau) = -\frac{3}{\ell^3} \delta(\vec{r}_1 -\vec{r}_2),
\label{difeq}
\end{eqnarray}
where $\alpha^2(\tau) = 3\tau/(2 \tau_0 \ell^2)$ with $\tau_0 = (4
k^2 D)^{-1}$. The solution of Eq.\ (\ref{difeq}) with Dirichlet
boundary conditions at $z=0$ and $z=L$ ($P=0$ if $z_1=0, L$ or
$z_2=0, L$) is readily found:\cite{skip97}
\begin{eqnarray}
P_0( \vec{r}_1, \vec{r}_2, \tau) &=& \frac{12 \pi}{\ell^3} \int
\mathrm{d}^2 \vec{p} \; \nonumber \\
&\times& \frac{\sinh[\beta_a(L-z_>)] \sinh(\beta_a z_<)}{\beta_a
\sinh(\beta_a L)}
\nonumber \\
&\times&\exp\left[ i(\vec{R}_1-\vec{R}_2)
\vec{p} \right].
\label{p0}
\end{eqnarray}
Here $\beta_a^2 = \vec{p}^2 + \alpha^2(\tau)$,
$z_>=\mathrm{max}\left\{z_1, z_2 \right\}$,
$z_<=\mathrm{min}\left\{z_1, z_2 \right\}$, and
the subscript ``0'' of $P_0$ denotes macroscopically
homogeneous case.
Inserting Eq.\ (\ref{p0}) into Eq.\ (\ref{ladder}), we
get
\begin{eqnarray}
C_1^{(0)}(\Delta \vec{q}_a, \Delta \vec{q}_b, \tau) =
\delta_{\Delta \vec{q}_a, \Delta \vec{q}_b}
\frac{3 \pi}{k^2 A}
\frac{\sinh^2(\beta_a \ell)}{\beta_a \ell \sinh(\beta_a L)}
\label{c10}
\end{eqnarray}
with $\beta_a^2 = \Delta \vec{q}_a^2 + \alpha^2(\tau)$. For
$\alpha^2(\tau)=0$, Eq.\ (\ref{c10}) reduces to the angular
correlation function,\cite{berk94} while for $\Delta\vec{q}_a =
\Delta\vec{q}_b = 0$ the time autocorrelation function of
transmitted light \cite{pine88} is recovered. The Kronecker delta
symbol in Eq.\ (\ref{c10}) describes the memory
effect.\cite{freund88}

Now we turn to the case of macroscopically heterogeneous medium,
assuming that the scatterer diffusion coefficient $D_{in}$ inside a
cylindrical region depicted in Fig.~\ref{fig1} is not the same as
$D_{out}$ in the surrounding medium (while $\ell$ is assumed to be
constant throughout the whole sample). The correlation function of
transmitted waves can be again described by Eqs.\ (\ref{ladder}),
(\ref{difeq}) but with $\alpha^2(\tau) = \alpha_0^2(\tau) +
\alpha_1^2(\tau) = 3\tau/(2\tau_{in} \ell^2)$ inside the inclusion
and $\alpha^2(\tau) = \alpha_0^2(\tau) = 3\tau/(2\tau_{out}
\ell^2)$ outside it. Here $\tau_{in, out} = (4 k^2 D_{in,
out})^{-1}$ and $\alpha_1^2(\tau) = 3\tau/(2 \ell^2) [1/\tau_{in} -
1/\tau_{out}]$.

\begin{figure}
\vspace{0.4cm} \psfig{file=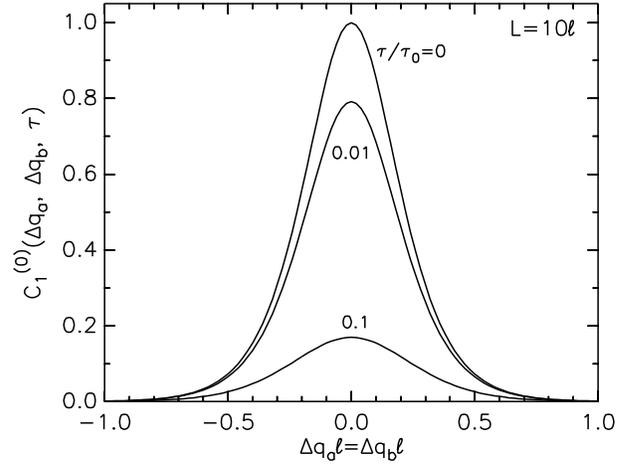,width=8cm} \vspace*{0.5cm}
\caption{Normalized angular-temporal correlation of a wave
transmitted through a macroscopically homogeneous (no inclusion)
slab of width $L= 10 \ell$ for three different time delays
$\tau/\tau_0 = 0, 0.01, 0.1$. This correlation function vanishes
identically for $\Delta \vec{q}_a \neq \Delta \vec{q}_b$, which
corresponds to the memory effect.\cite{freund88}} \label{fig2}
\end{figure}

Assuming $\left| \alpha_1^2(\tau) \right| \ll \alpha_0^2(\tau)$, we
can write an approximate solution of Eq.\ (\ref{difeq}) as a sum of
$P_0$ corresponding to the macroscopically homogeneous medium with
$\tau_0 = \tau_{out}$ [see Eq.\ (\ref{p0})], and $P_1$ which
describes the influence of inclusion:\cite{skip97}
\begin{eqnarray}
&&P_1(\left\{\vec{R}_1, \ell \right\}, \left\{\vec{R}_2, L-\ell \right\},
\tau) \simeq
\nonumber \\
&&\simeq -\alpha_1^2(\tau) \frac{\ell^3}{3} \int \mathrm{d}^3
\vec{r} \; P_0(\left\{\vec{R}_1, \ell \right\}, \vec{r}, \tau)
\nonumber \\
&&\mbox{~~~~~~~~~~~~~~~~~~} \times P_0(\vec{r}, \left\{\vec{R}_2,
L-\ell \right\}, \tau)
\nonumber \\
&&= -\frac{144 \pi^3}{3} \alpha_1^2(\tau) a^2 \frac{\Delta}{\ell}
\int\int \mathrm{d}^2 \vec{p} \; \mathrm{d}^2 \vec{s} \;
F_T(\vec{p}, \vec{s}, \tau)
\nonumber \\
&&\times \exp(i \vec{p} \vec{R}_1 - i \vec{s} \vec{R}_2 ),
\label{p1}
\end{eqnarray}
where the integration of the second line is taken over
the volume of inclusion, and $F_T$ is a form factor:
\begin{eqnarray}
&&F_T(\vec{p}, \vec{s}, \tau) = \frac{2 J_1\left( a \left| \vec{p}
- \vec{s} \right| \right)}{ a \left| \vec{p} - \vec{s} \right|}
\nonumber \\
&&\times \frac{\sinh(\beta_a \ell) \sinh(\beta_b \ell)}{
(\beta_a\ell) (\beta_b\ell) \sinh(\beta_a L) \sinh(\beta_b L)}
\nonumber \\
&&\times\left\{ \frac{\sinh\left[ (\beta_a - \beta_b) \Delta
\right]}{ (\beta_a - \beta_b) \Delta} \cosh\left[ \beta_a(L-z_0) +
\beta_b z_0 \right] \right.
\nonumber \\
&&-\left. \frac{\sinh\left[ (\beta_a + \beta_b) \Delta \right]}{
(\beta_a + \beta_b) \Delta} \cosh\left[ \beta_a(L-z_0) - \beta_b
z_0 \right] \right\}. \label{ft}
\end{eqnarray}
Here $J_1$ is the Bessel function of the first order, $\beta_a^2 =
\vec{p}^2 + \alpha_0^2(\tau)$, $\beta_b^2 = \vec{s}^2 +
\alpha_0^2(\tau)$, $\Delta = (z_2-z_1)/2$, and $z_0 = (z_1 +
z_2)/2$ (see Fig.~\ref{fig1}).

Inserting $P = P_0 + P_1$ into Eq.\ (\ref{ladder}), we obtain the
angular-temporal correlation function corresponding to the
macroscopically heterogeneous medium as a sum of two contributions:
$C_1 = C_1^{(0)} + C_1^{(1)}$, where $C_1^{(0)}$ is given by Eq.\
(\ref{c10}) with $\alpha(\tau) = \alpha_0(\tau)$, and
\begin{eqnarray}
C_1^{(1)}(\Delta \vec{q}_a, \Delta \vec{q}_b, \tau) &=& - 4 \pi
\alpha_1^2(\tau) \ell^2 \frac{3 \pi}{k^2 A} \nonumber \\
&\times& \frac{\pi a^2 }{A} \frac{\Delta}{\ell} F_T(\Delta
\vec{q}_a, \Delta \vec{q}_b, \tau). \label{c11}
\end{eqnarray}

\begin{figure}
\vspace{0.3cm} \psfig{file=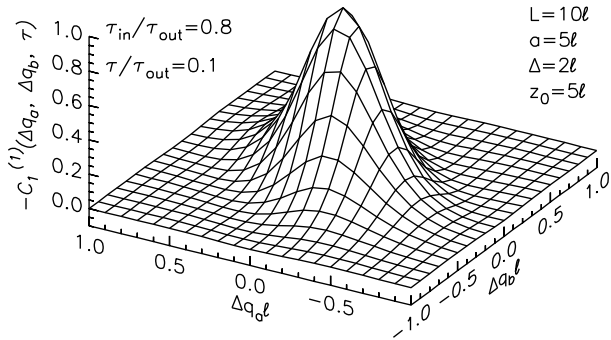,width=8cm} \vspace{0.5cm}
\caption{Normalized angular-temporal cross-correlation function
$C_1^{(1)}$ corresponding to the setup of Fig.~\ref{fig1}. We
assume $\Delta \vec{q}_a \parallel \Delta \vec{q}_b$ for this plot.
In contrast to $C_1^{(0)}$ (see Fig.~\ref{fig2}), $C_1^{(1)}$ is
not necessarily zero for $\Delta \vec{q}_a \neq \Delta \vec{q}_b$.
$C_1^{(1)} \neq 0$ only for $\tau \neq 0$.} \label{fig3}
\end{figure}

Equation (\ref{c11}) is the main result of the paper. We now
compare the $C_1^{(0)}$ correlation function, corresponding to a
macroscopically homogeneous slab [Eq.\ (\ref{c10})], and the
$C_1^{(1)}$ correlation function [Eq.\ (\ref{c11})], originating
from the presence of a dynamically heterogeneous region
(inclusion). As follows from Eq.\ (\ref{c10}), $C_1^{(0)}$ vanishes
identically if $\Delta \vec{q}_a \neq \Delta \vec{q}_b$, which is a
manifestation of the memory effect.\cite{freund88} If $\Delta
\vec{q}_a = \Delta \vec{q}_b$, $C_1^{(0)}$ decays to zero for
$\Delta q_a > 1/L$ (see Fig.~\ref{fig2}). In contrast, $C_1^{(1)}$
correlation is not necessarily zero for $\Delta \vec{q}_a \neq
\Delta \vec{q}_b$ (see Fig.~\ref{fig3}). The memory effect is still
present for the $C_1^{(1)}$ correlation function, as it is peaked
near $\Delta\vec{q}_a = \Delta\vec{q}_b$ due to $J_1( a \left|
\Delta\vec{q}_a - \Delta\vec{q}_b \right|)/ (a \left|
\Delta\vec{q}_a - \Delta\vec{q}_b \right|)$ term in
$F_T(\Delta\vec{q}_a, \Delta\vec{q}_b, \tau)$. The memory effect
for $C_1^{(1)}$ is considerably less sharp than for $C_1^{(0)}$, as
one can see from Fig.~\ref{fig3}.

\begin{figure}
\vspace{0.3cm} \psfig{file=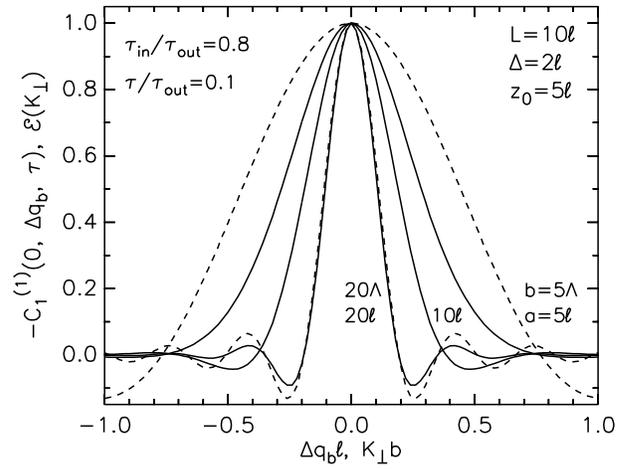,width=8cm} \vspace{0.5cm}
\caption{Normalized angular-temporal correlation functions of
transmitted waves for a single plane wave ($\Delta \vec{q}_a = 0$)
incident upon a slab with a cylindrical inclusion inside (solid
lines). Dashed lines show the (normalized) angular distribution of
the wave field ${\cal E}$ diffracted by a circular aperture of
radius $b$ ($b = 5\Lambda$ and $b = 20\Lambda$ with $\Lambda$ being
the wavelength). For $a = 20\ell \gg \Delta$ and $b=20\Lambda$,
correlation and diffraction curves are remarkably close, suggesting
that correlation is ``diffracted'' by the inclusion.} \label{fig4}
\end{figure}

Let us consider the simplest and practically important case of a
single incident plane wave ($\Delta \vec{q}_a = 0$). For a
macroscopically homogeneous slab, the angular-temporal
cross-correlation vanishes if $\Delta \vec{q}_b \neq 0$, i.e. the
waves scattered in different directions are
uncorrelated.\footnote{In reality, correlation persists as long as
$\Delta q_b < 1/W$, and our result (\ref{c10}) corresponds to the
limit $W \rightarrow \infty$.} If a heterogeneous region is
embedded inside the slab, the $C_1^{(1)}$ term appears and
correlation between the waves scattered along different directions
is not necessarily zero. The $C_1^{(1)}$ term is plotted in
Fig.~\ref{fig4} for three different radii $a$ of the inclusion
(solid lines). As is seen from the figure, the correlation range
can be estimated as $\Delta q_b \sim 1/a$. It is worthwhile to note
that the correlation between the waves scattered along different
directions ($\Delta\vec{q}_b \neq 0$), introduced by the inclusion,
exists only for $\tau \neq 0$. If $\tau = 0$, $C_1^{(1)} = 0$ and
the correlation function is given by $C_1^{(0)}$ which is
identically zero for $\Delta\vec{q}_b \neq \Delta\vec{q}_a = 0$.

Some qualitative insight into the behavior of the $C_1^{(1)}$
correlation can be gained by comparing Eq.\ (\ref{c11}) with the
angular distribution of the wave field ${\cal E}$ (wave number $K =
2\pi/\Lambda$) diffracted by a circular aperture of radius $b$ (see
Fig.~\ref{fig4}, dashed lines):\cite{born65}
\begin{eqnarray}
{\cal E}(K_{\perp}) \propto b^2 \frac{J_1(K_{\perp} b)}{K_{\perp} b},
\label{diffr}
\end{eqnarray}
where $K_{\perp}$ is the projection of $\vec{K}$ onto the plane of
the aperture. In the case of $a \gg \Delta$, which corresponds to a
``pill-shaped'' inclusion, the correlation function given by our
Eq.\ (\ref{c11}) and the diffraction pattern of Eq.\ (\ref{diffr})
are remarkably close (see, e.g., the curves corresponding to $a =
20\ell$ and $b=20\Lambda$ in Fig.~\ref{fig4}). In this case, one
can explain the appearance of correlation between the waves
scattered in different directions in a macroscopically
heterogeneous medium using the classical diffraction
theory,\cite{born65} and assuming $\Lambda = \ell$. As a
consequence, some theorems known for diffraction of waves (e.g.,
the Babinet's principle), apply directly to the angular-temporal
correlation function of light transmitted through a macroscopically
heterogeneous turbid medium. It is worthwhile to note that this
holds for any shape of inclusion, provided that the transverse
extent of inclusion is significantly greater than its extent along
the $z$-axis ($a \gg \Delta$ in our notation). For $a \sim \Delta$,
the quantitative agreement between Eqs.\ (\ref{c11}) and
(\ref{diffr}) is absent, although their overall behavior is similar
(see, e.g., the curves corresponding to $a = 5\ell$ and $b =
5\Lambda$ in Fig.~\ref{fig4}).

Up to now, our analysis has been devoted to the correlation
functions of transmitted light. In experiments, however, it could
be more convenient to work with diffusely reflected waves.
Calculation of the angular-temporal correlation function of
reflected waves is performed similarly to that of transmitted ones.
If $\left| \vec{q}_a + \vec{q}_b \right|$, $\left|
\vec{q}_a^{\prime} + \vec{q}_b^{\prime} \right|$, $\left| \vec{q}_a
+ \vec{q}_b^{\prime} \right|$, $\left| \vec{q}_a^{\prime} +
\vec{q}_b \right| \gg 1/\ell$, we can ignore the time-reversal
symmetry and obtain:
\begin{eqnarray}
C_1^{(0)}(\Delta \vec{q}_a, \Delta \vec{q}_b, \tau) &=&
\delta_{\Delta \vec{q}_a, \Delta \vec{q}_b} \frac{3 \pi}{k^2 A}
\nonumber \\
&\times& \frac{\sinh[\beta_a (L-\ell)] \sinh(\beta_a \ell)}{
\beta_a \ell \sinh(\beta_a L)}, \label{c10r}
\\
C_1^{(1)}(\Delta \vec{q}_a, \Delta \vec{q}_b, \tau) &=& - 4 \pi
\alpha_1^2(\tau) \ell^2 \frac{3 \pi}{k^2 A} \frac{\pi a^2 }{A}
\frac{\Delta}{\ell} \nonumber \\
&\times& F_R(\Delta \vec{q}_a, \Delta \vec{q}_b, \tau),
\mbox{~~~~~~} \label{c11r}
\end{eqnarray}
\begin{eqnarray}
&&F_R(\vec{p}, \vec{s}, \tau) = \frac{2 J_1\left( a \left| \vec{p}
- \vec{s} \right| \right)}{ a \left| \vec{p} - \vec{s} \right|}
\nonumber \\
&&\times \frac{\sinh(\beta_a \ell) \sinh(\beta_b \ell)}{
(\beta_a\ell) (\beta_b\ell) \sinh(\beta_a L) \sinh(\beta_b L)}
\nonumber \\
&&\times \left\{ \frac{\sinh\left[ (\beta_a + \beta_b) \Delta
\right]}{ (\beta_a + \beta_b) \Delta} \cosh\left[
(\beta_a+\beta_b)(L-z_0) \right] \right.
\nonumber \\
&&-\left. \frac{\sinh\left[ (\beta_a - \beta_b) \Delta \right]}{
(\beta_a - \beta_b) \Delta} \cosh\left[ (\beta_a-\beta_b)(L-z_0)
\right] \right\}. \label{fr}
\end{eqnarray}
The $C_1^{(0)}$ correlation function given by Eq.\ (\ref{c10r})
reduces to the result of Ref.\ \onlinecite{wang89} for $\tau = 0$,
$\beta_a L \rightarrow \infty$ and $\beta_a \ell \rightarrow 0$.
The $C_1^{(1)}$ term [Eq.\ (\ref{c11r})] has the same qualitative
features as the $C_1^{(1)}$ correlation of transmitted light [Eq.\
(\ref{c11})]. In reflection, however, low-order scattering events
become important for $\Delta q_a, \Delta q_b \sim 1/\ell$, and thus
our results (\ref{c10r})--(\ref{fr}) make sense only for $\Delta
q_a \ell, \Delta q_b \ell \ll 1$. If $\left| \vec{q}_a + \vec{q}_b
\right|$, $\left| \vec{q}_a^{\prime} + \vec{q}_b^{\prime} \right|$,
$\left| \vec{q}_a + \vec{q}_b^{\prime} \right|$, or $\left|
\vec{q}_a^{\prime} + \vec{q}_b \right|$ are of order of or smaller
than $1/\ell$, the time-reversal symmetry of the problem cannot be
ignored any more. This significantly complicates calculations even
in the case of macroscopically homogeneous medium.\cite{berk90}

In conclusion, we have calculated and discussed the angular-temporal
cross-correlation functions of waves scattered in a turbid,
dynamically heterogeneous medium. Our analysis demonstrates that the
considered correlation functions can be used to image a hidden
dynamic inclusion embedded in an otherwise homogeneous medium.
Comparison of our results with diffraction patterns obtained
for a wave diffracted by an aperture suggests that the
angular-temporal correlation function can be considered as
being ``diffracted'' by inclusion. Such an interpretation
is particularly successful for ``pill-shaped'' inclusions
which are much more extended in the transverse directions (i.e.,
in the directions parallel to the surfaces of the slab where they
are embedded) than in the longitudinal one.

\end{document}